\documentclass[11pt]{article}
\usepackage{amsmath, amssymb}
\usepackage{graphicx}

\numberwithin{equation}{section}

\allowdisplaybreaks

\newcommand{\eqa}{\begin{eqnarray}}
\newcommand{\eeqa}{\end{eqnarray}}
\newcommand{\beq}{\begin{equation}}
\newcommand{\eeq}{\end{equation}}

\newtheorem{dfn}{Definition}[section]
\newtheorem{thm}[dfn]{Theorem}

\newtheorem{lem}[dfn]{Lemma}
\newtheorem{cor}[dfn]{Corollary}
\newtheorem{emp}[dfn]{Example}

\def\res{\mathop{\mathrm {res}}\limits_}

\newcommand{\nn}{\nonumber}
\newcommand{\e}{\epsilon}

\newcommand{\p}{\partial}
\newcommand{\HH}{\mathcal{H}}
\newcommand{\bt}{\mathbf{t}}
\newcommand{\bs}{\mathbf{s}}
\newcommand{\bdzero}{\mathbf{0}}
\newcommand{\F}{\mathcal{F}}

\newcommand{\CC}{\mathbb{C}}
\newcommand{\Tr}{\mathrm{Tr}}
\newcommand{\be}{\begin{equation}}
\newcommand{\ee}{\end{equation}}
\newcommand{\M}{\mathcal{M}}

\newenvironment{prf}{\noindent {\it Proof} \ }{\hfill $\Box$}

\newenvironment{prfn}[1]{\noindent {\it Proof of #1} \ }{\hfill $\Box$}

\usepackage{footmisc}
\DefineFNsymbols{mySymbols}{{\ensuremath\dagger}{\ensuremath\ddagger}\S\P
   *{**}{\ensuremath{\dagger\dagger}}{\ensuremath{\ddagger\ddagger}}}
\setfnsymbol{mySymbols}
\newcommand{\indicationfootnote}{\thanks}

\begin{document}
\title{Remarks on intersection numbers and integrable \\ hierarchies. I. Quasi-triviality}
\author{Boris Dubrovin\indicationfootnote{Deceased on March 19, 2019.}\,, \quad Di Yang}
\date{}
\maketitle

\begin{abstract}
Explicit expression for quasi-triviality of scalar non-linear PDE is under consideration.
\end{abstract}

\section{Introduction}
Recall that the Korteweg--de Vries (KdV) equation
\be\label{KdV-eqn}
u_t = u u_x + \frac{\e^2}{12}u_{xxx}
\ee
and the modified KdV (mKdV) equation 
\be\label{mKdV}
w_t = \frac {w^2}{2}  w_x + \frac{\e^2}{12} w_{xxx}
\ee
are related via a Miura transformation~\cite{Miura}
\be\label{Miura}
u=\frac12 w^2+\sqrt{-1}\,\frac{\e}{2} w_x.
\ee
This means that, substituting~\eqref{Miura} and its unique formal inverse
\be
w = (2u)^{\frac12} -\sqrt{-1} \, \frac{\e}{4} \frac{ (2u)_x}{ 2u} + 
\e^2\biggl(\frac{5}{32}\frac{ (2u)_x^2}{ (2u)^{5/2}}-\frac{1}{8} \frac{(2u)_{xx}}{(2u)^{3/2}}\biggr)+\mathcal{O}(\e^3)
\ee
in~\eqref{KdV-eqn}, one gets~\eqref{mKdV}; {\it vice versa}.

The KdV and mKdV equations are examples of scalar evolutionary PDEs of the form~\cite{DZ-norm}:
\vspace{-3mm}
\begin{subequations}\label{normal-form}
\begin{align}
 & u_t=f(u) u_x+\e\bigl[a_1(u) u_{xx}+a_2(u) u_x^2\bigr] 
 + \e^2\bigl[a_3(u) u_{xxx} + a_4(u) u_{xx}u_x + a_5(u) u_x^3\bigr]
 + \cdots,\\
 & f(u)\not\equiv 0.
 \end{align}
\end{subequations}
Here, $\e$ is a parameter, and $f(u)$, $a_1(u)$, $a_2(u)$, $\cdots$ are given smooth functions of~$u$. 
Following \cite{DZ-norm,LZ}, we say that a change of the dependent variable of the form
\be\label{Miu}
w=W(u)+\sum_{k\geq 1} \e^k W^{[k]}(u; u_x,\dots,u_k)
\ee
is a Miura type transformation, if $W'(u)\not\equiv 0$ and $W^{[k]}$ is  
a degree~$k$ homogeneous differential polynomial of~$u$. 
Here, $u_j:=\p_x^j(u)$ 
is assigned the degree: $\deg u_j=j$, $j\geq 0$. 
All Miura type transformations form the Miura group.

The $\e\rightarrow 0$ limit of equation~\eqref{normal-form}
\be
v_t=f(v) v_x,\quad f(v)\not\equiv 0
\ee
is an evolutionary PDE of hydrodynamic type~\cite{DN}. The simplest non-trivial example is the 
dispersionless KdV equation ({\it aka} the Riemann--Hopf equation or the inviscid Burgers equation):
\be\label{dis-kdv}
v_t=v v_x.
\ee
This equation is NOT equivalent to
the KdV equation~\eqref{KdV-eqn} with $\e\neq 0$ under the Miura group action.
However, there exists a remarkable invertible transformation \cite{BGI,DZ-norm,LZ}
\be
u=v+\e^2\p_x^2 \biggl(\frac1{24}\log v_x\biggr)
+\e^4 \p_x^2\biggl(\frac{v_{xxxx}}{1152 v_x^2}-\frac{7v_{xx}v_{xxx}}{1920v_x^3}+\frac{v_{xx}^3}{360 v_x^4}\biggr)+\mathcal{O}\bigl(\e^6\bigr)
\ee
transforming~\eqref{KdV-eqn} to~\eqref{dis-kdv}. 
Such a transformation is called a {\it quasi-Miura transformation} \cite{DZ-norm,LZ}. 
We mention that the difference between Miura type and quasi-Miura transformations is simply 
that the latter allows rational and logarithmic dependence in~$v_x$. We also mention that 
$\p_x= \sum_{j\geq 0} v_{j+1} \p_{v_j} =  \sum_{j\geq 0} u_{j+1} \p_{u_j}$.

A scalar evolutionary PDE~\eqref{normal-form} is called {\it quasi-trivial} or say {\it possessing quasi-triviality}, if it can 
be transformed to its $\e\rightarrow 0$ limit via a quasi-Miura transformation.

\smallskip

\noindent \textbf{Liu--Zhang's Theorem (\cite{LZ}).}
{\em For an evolutionary PDE of the form~\eqref{normal-form} with $f'(u)\not\equiv0$, 
there exists a unique (under some homogeneity condition; see Theorem 4.3 of~\cite{LZ}) 
quasi-Miura transformation 
reducing~\eqref{normal-form} to its dispersionless limit.}

\smallskip

In this article we consider the following problem.

\smallskip

\noindent {\bf Problem~A}.  {\it Give an explicit expression of quasi-triviality of the KdV equation~\eqref{KdV-eqn}.}

\smallskip

\noindent This problem is {\it algebraic}, but the solution turns out to be {\it topological}.
 
Before presenting a solution to Problem~A, we recall some standard
notations. For $j\geq 0$, denote $v^{(j)}=v_j:= \p_x^j (v)$. 
By a partition $\lambda$, we mean a non-increasing sequence of non-negative integers 
$(\lambda_1, \dots, \lambda_{\ell(\lambda)},0,\dots)$, where $\ell(\lambda)$ 
denotes the length of $\lambda$, and $\lambda_1\geq \lambda_2\geq \dots\geq \lambda_{\ell(\lambda)}$ 
are the nonzero components of~$\lambda$.
The set of all partitions is denoted by $\mathbb{Y}$.  
Denote by $|\lambda|:=\sum_{j=1}^{\ell(\lambda)} \lambda_j$ the weight of~$\lambda$,  
by $\mathbb{Y}_k$ the set of partitions of weight~$k$, 
and by $m_i(\lambda)$ the multiplicity of~$i$ in~$\lambda$, $i\geq 1$. Denote also
\be\label{standard}
m(\lambda)!=\prod_{i\geq 1} m_i(\lambda)!,\qquad z_\lambda=m(\lambda)! \, \prod_{i\geq 1} i^{m_i(\lambda)}.
\ee
The partition of~$0$ is denoted by $(0)$, with $\ell((0)):=0$ and $|(0)|:=0$.  For any $\lambda,\mu\in\mathbb{Y}$, define $\lambda+\mu:=(\lambda_1+\mu_1, \lambda_2+\mu_2,\dots)$. Define
$\lambda+1:=\lambda+\bigl(1^{\ell(\lambda)}\bigr)$ if $\lambda\neq (0)$, and $(0)+1:=(0)$.
For an arbitrary sequence of indeterminates $q_1,q_2,\dots$, denote $q_\lambda:=q_{\lambda_1}\cdots q_{\lambda_{\ell(\lambda)}}$ if $\lambda\neq(0)$, and $q_{(0)}:=1$.  
Introduce also some integers:  For any $\lambda,\mu\in\mathbb{Y}$, define
\be
Q^{\lambda\mu}:=(-1)^{\ell(\lambda)}\sum_{\substack{\mu^1\in \mathbb{Y}_{\lambda_1},\dots,\mu^{\ell(\lambda)}\in\mathbb{Y}_{\lambda_{\ell(\lambda)}} \\ \sum_{q=1}^{\ell(\lambda)} \mu^q=\mu}} \prod_{q=1}^{\ell(\lambda)}
\frac{\left(\lambda_q+\ell(\mu^q)\right)!\,(-1)^{\ell(\mu^q)}}{m(\mu^q)! \prod_{j=1}^\infty (j+1)!^{m_j(\mu^q)}}.
\ee
We note that $Q^{\lambda\mu}=0$ unless $|\lambda|=|\mu|$, and call $\bigl(Q^{\lambda\mu}\bigr)_{|\lambda|=|\mu|}$ the {\it $Q$-matrices}.

\begin{thm}\label{quasi-KdV} 
Quasi-triviality of the KdV equation \eqref{KdV-eqn} has the following expression:
\begin{align}
& u=v+ \e^2  \p_x^2 \biggl(M_1(v_x) + \sum_{g=2}^\infty \e^{2g-2} M_g(v_x,v_{xx},\dots,v_{3g-2})\biggr),\\
& M_1(v_x)=\frac1{24} \log v_x, \label{M1-kdv}\\
& M_g(v_x,v_{xx},\dots,v_{3g-2})=\sum_{\lambda,\mu\in\mathbb{Y}_{3g-3}} \frac{\langle \tau_{\lambda+1} \rangle}{m(\lambda)!} \; Q^{\lambda\mu}\, \frac{v_{\mu+1}}{v_1^{\ell(\mu)+g-1}}, \qquad g\geq 2. \label{M-g}
\end{align}
Here, $\langle\tau_{\lambda+1}\rangle$ are the intersection numbers of $\psi$-classes on the Deligne--Mumford moduli spaces (for 
the definitions of these numbers see equation~\eqref{Hodgeintegralsdef}). 
\end{thm}
The proof is in Section~\ref{proof}.

The following three more problems will also be considered. 

\smallskip

\noindent {\bf Problem B.} {\it Give an explicit quasi-triviality of the intermediate Long wave (ILW) equation
\be 
u_t=u u_x+ \sum_{g\ge1}\e^{2g}s^{g-1}\frac{|B_{2g}|}{(2g)!}u_{2g+1}.
\ee
Here $B_{2g}$ denote the Bernoulli numbers, defined by
$\frac{x}{1-e^{-x}} =: \sum_{n=0}^\infty \frac{B_k}{k!} x^k$.}

\smallskip

\noindent {\bf Problem C.} {\it Give an explicit quasi-triviality of the discrete KdV equation}
\be u_t=\frac{1}{\e}\left(e^{u(x+\e)}-e^{u(x-\e)}\right).\ee 

\smallskip

\noindent {\bf Problem D.} 
{\it Give an explicit quasi-triviality of the Burgers equation}
\beq\label{Burgerseqintro}
u_t= u u_x+ \e u_{xx}.
\eeq

We remark that for an integrable PDE of the form~\eqref{normal-form} with $f'(u)\not\equiv 0$,
the quasi-triviality of this PDE is always the property of the whole corresponding integrable hierarchy.

We will see from solutions to the above problems in Sections~\ref{proof}--\ref{Burgers} that 
 the associated essential numbers to each problem (the primitive Hodge integrals 
 for the case of Problems A,B,C, and enumeration of graphs with 
 valencies~$\geq 3$ for the case of Problem D) 
are {\it all} contained in a simple nonlinear equation (KdV, ILW, discrete KdV, 
Burgers, respectively); this is revealed by Liu--Zhang's theorem and by the so-called inverse Q-matrices 
(see Definition~\ref{inv-Q-nm} below).

\paragraph{Organization of the paper.} 
In Section~\ref{sect2} we review the topological solution 
to the Riemann hierarchy. In Section~\ref{Hodge} we review 
the construction of Hodge hierarchy. 
In Sections~\ref{proof}--\ref{Burgers} 
we give solutions to Problems A--D. Concluding remarks are 
given in Section~\ref{further-rmks}. 
Straightforward proof 
of a technical lemma (Lemma~\ref{thm-dlyz}) is given in Appendix~\ref{app1014}.

\paragraph{Acknowledgements.}
D.Y. is grateful to Youjin Zhang for his advising.
Part of the work of D.Y. was done while he was a post-doc at SISSA; 
he thanks SISSA for excellent working conditions and 
financial supports. 

\section{Riemann hierarchy and $Q$-matrices}
\label{sect2}
The goal of this section is to do some preparations for the later sections.
Recall that an evolutionary PDE of the form
\beq
u_s=g(u) u_x+\e \bigl[b_1(u) u_{xx}+b_2(u) u_x^2 \bigr] +\e^2 \bigl[b_3(u) u_{xxx} + b_3(u) u_{xx}u_x + b_4(u) u_x^3\bigr]+\cdots
\eeq
is called an infinitesimal symmetry of~\eqref{normal-form} if  
\[\p_s  \p_t u=\p_t  \p_s  u.\]
Following~\cite{Dubrovin}, we say 
equation~\eqref{normal-form} with $f'(u)\not\equiv 0$ is called {\it integrable}, if it possesses an infinite family of infinitesimal 
symmetries parameterized by a smooth function of one variable.

The Riemann--Hopf equation \eqref{dis-kdv} is integrable: for any smooth function $g(v)$, the PDE
\be\label{infsymm}
v_s=g(v) v_x
\ee
gives an infinitesimal symmetry of~\eqref{dis-kdv}. 
Let us look at a particular sub-family in these infinitesimal symmetries
\be\label{riemann}
v_{t_k}=\frac{v^k}{k!}v_x,\qquad k\geq 0.
\ee
Observe that equations~\eqref{riemann} commute pairwise, i.e., 
\be\label{commpairwise}
\p_{t_i} \p_{t_j} v= \p_{t_j} \p_{t_i} v, \qquad \forall\, i,j\geq 0.
\ee
Thus \eqref{riemann} can be solved together. They are called the {\em Riemann hierarchy}. 
The $k=1$ equation in~\eqref{riemann} is the Riemann--Hopf equation~\eqref{dis-kdv}. The $k=0$ equation
reads $u_{t_0}=u_x$, so we identify~$t_0$ with~$x$.

We will be interested in solutions of~\eqref{riemann} in the formal power series ring $\mathbb{C}[[\bt]]$. 
Indeed, consider the initial value problem of~\eqref{riemann} along with the initial condition:
\be\label{ivp-f0}
v(x,0,0,\dots) =f_0(x) \in \mathbb{C}[[x]],\quad f_0'(0)\neq 0.
\ee 
Here, the condition $f_0'(0)\neq 0$ implies $f_0(x)$ has a compositional inverse in $\mathbb{C}[[x]]$. 
Denote by $c_p$, $p\geq 0$ the Taylor coefficients of $f^{-1}_0(x)$, i.e., 
\be
\sum_{p\geq 0} \frac{c_p}{p!} x^p=f_0^{-1}(x),\quad c_1\neq 0.
\ee
As explained above, equations~\eqref{riemann} have a unique solution~$v(\bt)$ in $\mathbb{C}[[\bt]]$
satisfying~\eqref{ivp-f0}. 
\begin{lem}[\cite{DZ-norm}] \label{el-ri}
The unique solution $v(\bt)$ satisfies the following equation:
\be\label{el}
\tilde t_0+\sum_{p\geq 1} \tilde t_p \frac{v(\bt)^p}{p!}=0,
\ee
where $\tilde t_p:=t_p-c_p$ ($c_1\neq 0$). Moreover, solution to~\eqref{el} in~$\mathbb{C}[[\bt]]$ is unique.
\end{lem}
Equation~\eqref{el} is called the genus zero Euler--Lagrange equation 
for the KdV hierarchy~\cite{Du1,DZ-norm}.

We now focus on a particular solution to the Riemann 
hierarchy~\eqref{riemann}, denoted by~$v^{\rm top}(\bt)$, that is 
specified by the initial data $f_0(x)=x$. 
In other words, $c_p=\delta_{p,1}$ is under consideration. 
The $v^{\rm top}(\bt)$ is often called the {\it topological solution}.
\begin{lem}\label{lem-vtop}
The $v^{\rm top}(\bt)$ has the explicit expression
\be\label{vtop}
v^{\rm top}(\bt)=\sum_{k\geq 1} \frac{1}{k}\sum_{p_1,\dots, p_k\geq 0 \atop p_1+\cdots+p_k=k-1} \frac{t_{p_1}}{p_1!}\cdots\frac{t_{p_k}}{p_k!}.
\ee
\end{lem}
\begin{prf} For any solution $v(\bt)$ in $\mathbb{C}[[\bt]]$ to the Riemann hierarchy~\eqref{riemann}, we have
\begin{align}
& v_{t_{k_1}}=\p_x \left(\frac{v^{k_1+1}}{k_1!\,(k_1+1)}\right),\\
& v_{t_{k_1} t_{k_2}}=\p_x^2 \left(\frac{v^{k_1+k_2+1}}{k_1! \, k_2! \, (k_1+k_2+1)}\right),\\
& v_{t_{k_1} \dots t_{k_N}}=\p_x^N \left(\frac{v^{k_1+\dots+k_N+1}}{k_1! \cdots k_N!\, (k_1+\dots +k_N+1)}\right), \quad \forall\,N\geq 3.
\end{align}
Here $k_1,k_2$, \dots, $\geq 0$. The lemma is then proved by noticing $v^{\rm top}_k (\bdzero)=\delta_{k,1}$.
\end{prf}

Proceed with a simplification of~\eqref{vtop}. 
Applying $\p_x^m$ on the both sides of~\eqref{vtop} we obtain
\be\label{top-jet-expression}
v_m^{\rm top}(\bt)=\sum_{k\geq 1} \sum_{p_1,\dots,p_k\geq 0 \atop p_1+\cdots+p_k=k+m-1} (k+1)\cdots(k+m-1)\, 
\frac{t_{p_1}}{p_1!}\cdots\frac{t_{p_k}}{p_k!},\qquad m\geq 1,
\ee
where we recall that $v^{\rm top}_m(\bt):=\p_x^m\bigl(v^{\rm top}(\bt)\bigr)$. The following shorthand notations will be used:
\begin{itemize}
\item[(i)] Denote $v(\bt)=v^{\rm top}(\bt)$ unless otherwise specified.
\item[(ii)] Denote $v^s=v^s(t_1,t_2,\dots):=v(\bt)|_{t_0=0}$,  and denote $v_j^s=v_j^s(t_1,t_2,\dots):=v_j(\bt)|_{t_0=0}$, $j\geq 1$.
\end{itemize}
Obviously $v^s=0$. The Taylor expansion of~$v(\bt)$ with respect to~$x$ then reads 
\be
v(\bt)=v^s+\sum_{j\geq 1} v_j^s \, \frac{x^j}{j!}=\sum_{j\geq 1} v_j^s \, \frac{x^j}{j!}.
\ee
\begin{lem} \label{Lagrange} 
For $m\geq 1$, the following formula holds true:
\begin{align}
v_m^s 
&= \sum_{\mu\in\mathbb{Y}_{m-1}} 
\frac{\left(m-1+\ell(\mu)\right)!}{\prod_{j\geq 1} (j+1)!^{m_j(\mu)}} \,\frac{t_{\mu+1}}{m(\mu)! \, (1-t_1)^{m+\ell(\mu)}}. \label{vn}
\end{align}
\end{lem}
\begin{prf} 
For $m=1$, we know from~\eqref{top-jet-expression} that $v_1^s=1/(1-t_1)$.
For $m\geq 2$,
\begin{align}
& v_m^s 
= \sum_{k\geq 1} \sum_{\lambda\in\mathbb{Y}_{k+m-1} \atop \ell(\lambda)=k} 
\frac{(k+m-1)!}{k!} \, \binom{k}{m_1(\lambda),m_2(\lambda),\cdots} \, \frac{t_\lambda}{\prod_{i\geq 1} i!^{m_i(\lambda)}} \nn\\
& \quad = \sum_{m_1,m_2,\dots \geq 0\atop \sum_{j\geq 1} (j-1) m_j=m-1} \biggl(m_1+\sum_{j\geq 2} jm_j\biggr)! \, 
\prod_{j\geq 1} \frac{t_j^{m_j}}{ j!^{m_j}  m_j!}\nn\\
& \quad = \sum_{m_2,m_3,\dots \geq 0\atop \sum_{j\geq 2} (j-1) m_j=m-1} \biggl(\sum_{j\geq 2} jm_j\biggr)! \, 
\prod_{j\geq 2} \frac{t_j^{m_j}}{ j!^{m_j}  m_j!} \frac{1}{(1-t_1)^{1+\sum_{j\geq 2} j m_j}} , \nn
\end{align}
where the last equality uses Newton's binomial identity:
$(1-x)^{-1-k}=\sum_{s\geq 0} \binom{s+k}{k} x^s$.
\end{prf}

For each partition $\mu\in\mathbb{Y}$, we call the integer
\be\label{lndef}
L(\mu):=\frac{\bigl(|\mu|+\ell(\mu)\bigr)!\, (-1)^{\ell(\mu)}}{m(\mu)! \, \prod_{j\geq 1} (j+1)!^{m_j(\mu)}}=(-1)^{\ell(\mu)} \frac{|\mu+1|!}{z_{\mu+1}}
\ee
the {\it Lagrange number} associated to~$\mu$.
The first few Lagrange numbers are 
\[ L((0))=1, \quad
L((1))=-1,\quad  L((2))=-1, \quad L\left(\bigl(1^2\bigr)\right)=3, \quad L((3))=-6,\]
\[L((2,1))=10,\quad L\left(\bigl(1^3\bigr)\right)=-15, \quad L((n))=-1, \quad L\left(\bigl(1^n\bigr)\right)=(-1)^n (2n-1)!!.\]
Using the Lagrange number we can write formula~\eqref{vn} as
\be\label{vn-short}
v_n^s = \sum_{\mu\in\mathbb{Y}_{n-1}} 
(-1)^{\ell(\mu)} \, L(\mu) \,\frac{t_{\mu+1}}{(1-t_1)^{n+\ell(\mu)}},\qquad n\geq 1.
\ee

\begin{lem} [Zhou~\cite{Zhou}] \label{Zhou-lemma} 
The following formulae hold true:
\eqa
&& 1-t_1=\frac{1}{v_1^s},\label{t1}\\
&& -t_k=\sum_{\mu\in \mathbb{Y}_{k-1}}  L(\mu) \, \frac{v_{\mu+1}^s}{(v_1^s)^{1+|\mu+1|}},\quad k\geq 2.\label{Zhou}
\eeqa
\end{lem}

We remark that both~\eqref{vn-short} and Lemma~\ref{Zhou-lemma} can also be proved straightforwardly 
by using the Lagrange inversion (cf. e.g.~\cite{L,J}). 
\begin{dfn} \label{Q-nm} For any two partitions $\lambda,\mu$, define $Q^{(0)(0)}:=1$, and define 
\be
Q^{\lambda\mu}:=(-1)^{\ell(\lambda)}\sum_{\substack{\mu^1\in \mathbb{Y}_{\lambda_1},\dots,\mu^{\ell(\lambda)}\in\mathbb{Y}_{\lambda_{\ell(\lambda)}} \\ \mu^1\cup \mu^2\cup \cdots\cup \mu^{\ell(\lambda)} =\mu}} \prod_{q=1}^{\ell(\lambda)}L(\mu^q).
\ee 
For $k\geq 0$, we call $\bigl(Q^{\lambda\mu}\bigr)_{|\lambda|=|\mu|=k}$ the $Q$-matrices.
\end{dfn}
\begin{lem}\label{tv-lem}
The following formula holds true:
\be\label{tv}
t_{\lambda+1}= \sum_{\mu\in\mathbb{Y}_{|\lambda|}} \, Q^{\lambda\mu}\,\frac{v^s_{\mu+1}}{(v_x^s)^{l(\mu)+|\lambda+1|}},\qquad \forall\,\lambda\in\mathbb{Y}.
\ee
\end{lem}
\begin{prf} 
For any partition $\lambda\in\mathbb{Y}$, we have
\begin{align}
t_{\lambda+1}&= (-1)^{\ell(\lambda)}\, \prod_{q=1}^{l(\lambda)} \sum_{\mu\in \mathbb{Y}_{\lambda_q}} 
L(\mu) \frac{v^s_{\mu+1}}{ (v_x^s)^{1+|\mu+1|}} \nn\\
&= (-1)^{\ell(\lambda)}\, \sum_{\mu^1\in \mathbb{Y}_{\lambda_1},\dots,\mu^{\ell(\lambda)}\in\mathbb{Y}_{\lambda_{\ell(\lambda)}}} \prod_{q=1}^{\ell(\lambda)}
L(\mu^q) \,\frac{v^s_{\mu^q+1}}{(v_x^s)^{l(\mu^q)+\lambda_q+1}} \nn\\
&= \sum_{\mu\in\mathbb{Y}_{|\lambda|}} \,\frac{v^s_{\mu+1}}{(v_x^s)^{l(\mu)+|\lambda+1|}}\, (-1)^{\ell(\lambda)} \sum_{\substack{\mu^1\in \mathbb{Y}_{\lambda_1},\dots,\mu^{l(\lambda)}\in\mathbb{Y}_{\lambda_{l(\lambda)}} \\ \mu^1\cup \mu^2\cup \cdots\cup \mu^{\ell(\lambda)} =\mu}} L(\mu^q). \nn
\end{align}
The lemma is proved.
\end{prf}

\begin{dfn}\label{inv-Q-nm}  Define $Q_{(0)(0)}:=1$, and define
\be
Q_{\mu\rho}:=\sum_{\substack{\rho^1\in \mathbb{Y}_{\mu_1},\dots,\rho^{\ell(\mu)}\in\mathbb{Y}_{\mu_{\ell(\mu)}} \\ \rho^1\cup \rho^2\cup \cdots\cup \rho^{\ell(\mu)} =\rho}} \prod_{q=1}^{\ell(\mu)} |L(\rho^q)|,
\ee 
where $\mu,\rho$ are two arbitrary partitions. We call $(Q_{\mu\rho})_{|\mu|=|\rho|}$ the inverse $Q$-matrices.
\end{dfn}

\begin{lem}
The following formula holds true:
\be
v_{\mu+1}^s= \sum_{\rho\in\mathbb{Y}_{|\mu|}} \,Q_{\mu\rho}\,\frac{t_{\rho+1}}{(1-t_1)^{\ell(\rho)+|\mu+1|}},  \qquad \forall\,\mu\in\mathbb{Y}.
\ee
\end{lem}
\begin{prf} For any partition $\mu\in\mathbb{Y}$, we have
\begin{align}
v_{\mu+1}&= \prod_{q=1}^{l(\mu)} \sum_{\rho\in \mathbb{Y}_{\mu_q}} 
|L(\rho)| \frac{t_{\rho+1}}{ (1-t_1)^{1+|\rho+1|}} \nn\\
&= \sum_{\rho^1\in \mathbb{Y}_{\mu_1},\dots,\rho^{\ell(\mu)}\in\mathbb{Y}_{\mu_{\ell(\mu)}}} \prod_{q=1}^{\ell(\mu)}
|L(\rho^q)| \,\frac{t_{\rho^q+1}}{(1-t_1)^{l(\rho^q)+\mu_q+1}} \nn\\
&=
\sum_{\rho\in\mathbb{Y}_{|\mu|}} \,\frac{t_{\rho+1}}{(1-t_1)^{\ell(\rho)+|\mu+1|}} 
\sum_{\substack{\rho^1\in \mathbb{Y}_{\mu_1},\dots,\rho^{\ell(\mu)}\in\mathbb{Y}_{\mu_{\ell(\mu)}} \\ \rho^1\cup \rho^2\cup \cdots\cup \rho^{\ell(\mu)} =\rho}} |L(\rho^q)|. \nn
\end{align}
The lemma is proved.
\end{prf}

\begin{lem}\label{QQ-pro} 
We have
\begin{itemize}
\item[a)] ~ $Q^{\lambda\mu}=Q_{\lambda\mu}=0$ if $|\lambda|\neq |\mu|$.
\item[b)] ~ The $Q$-matrices $\bigl(Q^{\lambda\mu}\bigr)_{|\lambda|=|\mu|}$ and the inverse $Q$-matrices $(Q_{\lambda\mu})_{|\lambda|=|\mu|}$ are upper triangular with respect to the reverse lexicographic ordering.
\item[c)] ~ $Q^{\lambda\mu}$ are integers and $Q_{\lambda\mu}$ are positive integers.
\item[d)] ~ $Q^{\lambda\lambda}=1,\,Q_{\lambda\lambda}=1,\quad \forall\,\lambda\in\mathbb{Y}$.
\item[e)] ~ $Q^{(n)\,\mu}=-L(\mu),\quad Q_{(n)\,\mu}=|L(\mu)|,\,\quad\forall\mu\in\mathbb{Y}$.
%\item[f).] $Q_{\lambda\,(1^{|\lambda|})}=A134145 in oeis$.
\item[f)] ~ $\forall\,k\geq 0$, $(Q_{\lambda\mu})_{|\lambda|=|\mu|=k} \, \bigl(Q^{\lambda\mu}\bigr)_{|\lambda|=|\mu|=k}
=I$, where $I$ denotes the identity matrix.
\end{itemize}
\end{lem}
\begin{prf}
a)--e) are easy consequences of 
Definition\,\ref{Q-nm} and~\eqref{lndef}. Note that $\forall\,\lambda,\rho\in\mathbb{Y}$,
\be
t_{\lambda+1}=\sum_{\mu\in\mathbb{Y}_{|\lambda|}} \, Q^{\lambda\mu}\,\frac{v^s_{\mu+1}}{(v_x^s)^{l(\mu)+|\lambda|+1}}
=\sum_{\mu,\rho\in\mathbb{Y}_{|\lambda|}} \, Q^{\lambda\mu}\,
Q_{\mu\rho}\frac{1}{(v_x^s)^{l(\mu)+|\lambda+1|}}\frac{t_{\rho+1}}{(1-t_1)^{l(\rho)+|\mu+1|}}.
\ee
The assertion~f) is then proved by noticing that $1-t_1=\frac{1}{v_1^s}$.
\end{prf}

The first several $Q$-matrices and inverse $Q$-matrices are given by
\begin{align}
&(Q^{\lambda\mu})=(1),\quad (Q_{\lambda\mu})=(1), \quad |\lambda|=|\mu|=0; \nn \\
&(Q^{\lambda\mu})=(1),\quad (Q_{\lambda\mu})=(1), \quad |\lambda|=|\mu|=1; \nn \\
&(Q^{\lambda\mu})=
\begin{pmatrix} 
1 & -3\\ 0 & 1\\ 
\end{pmatrix}, \quad (Q_{\lambda\mu})=\begin{pmatrix} 1 & 3\\ 0 & 1\\ \end{pmatrix},
\quad |\lambda|=|\mu|=2; \nn \\
&(Q^{\lambda\mu})=\begin{pmatrix} 
1 & -10 & 15 \\ 0 & 1 & -3\\ 0 & 0& 1\\ \end{pmatrix}, 
\quad (Q_{\lambda\mu})=
\begin{pmatrix} 
1 & 10 & 15 \\ 0 & 1 & 3\\ 0 & 0& 1\\ 
\end{pmatrix},\quad |\lambda|=|\mu|=3; \nn \\
&(Q^{\lambda\mu})={\footnotesize \begin{pmatrix} 
1 & -15 & -10 & 105 & -105\\ 0 & 1 & 0 & -10 & 15 \\ 0 & 0& 1 & -6 & 9\\ 0&0&0&1&-3\\0&0&0&0&1 \end{pmatrix}}, 
\quad (Q_{\lambda\mu})= {\footnotesize
\begin{pmatrix} 
1 & 15 & 10 & 105 & 105\\ 0 & 1 & 0 & 10 & 15 \\ 0 & 0& 1 & 6 & 9\\ 0&0&0&1&3\\0&0&0&0&1 
\end{pmatrix}}, \quad |\lambda|=|\mu|=4; \nn \\
&(Q^{\lambda\mu})=
{\scriptsize \begin{pmatrix}
 1 & -21 & -35 & 210 & 280 & -1260 & 945 \\
 0 & 1 & 0 & -15 & -10 & 105 & -105 \\
 0 & 0 & 1 & -3 & -10 & 45 & -45 \\
 0 & 0 & 0 & 1 & 0 & -10 & 15 \\
 0 & 0 & 0 & 0 & 1 & -6 & 9 \\
 0 & 0 & 0 & 0 & 0 & 1 & -3 \\
 0 & 0 & 0 & 0 & 0 & 0 & 1
 \end{pmatrix}}, ~ (Q_{\lambda\mu})={\scriptsize \begin{pmatrix}
 1 & 21 & 35 & 210 & 280 & 1260 & 945 \\
 0 & 1 & 0 & 15 & 10 & 105 & 105 \\
 0 & 0 & 1 & 3 & 10 & 45 & 45 \\
 0 & 0 & 0 & 1 & 0 & 10 & 15 \\
 0 & 0 & 0 & 0 & 1 & 6 & 9 \\
 0 & 0 & 0 & 0 & 0 & 1 & 3 \\
 0 & 0 & 0 & 0 & 0 & 0 & 1 \\
\end{pmatrix}},\nn\\
& \qquad \qquad \qquad \qquad\qquad \qquad \qquad \qquad \qquad\qquad\qquad\qquad\qquad\qquad\qquad\qquad\qquad |\lambda|=|\mu|=5. \nn
\end{align}

\section{Hodge integrals and integrable hierarchies: a short review}\label{Hodge}
Let $\overline{\M}_{g,n}$ be the Deligne--Mumford moduli space of 
stable algebraic curves of genus~$g$ with~$n$ distinct marked points. 
Here the non-negative integers $g,n$ should satisfy the stability condition
\be\label{stability}
2g-2+n>0.
\ee
Denote by $\mathcal{L}_i$, $i=1,...,n$ the $i_{\text{th}}$ tautological line bundle on $\overline{\M}_{g,n}$, 
by $\mathbb{E}_{g,n}$ the Hodge bundle on $\overline{\M}_{g,n}$, 
by $\psi_i:=c_1(\mathcal{L}_i)$ the $\psi$-class,  
and by $\lambda_j:=c_j(\mathbb{E}_{g,n})$, $j=1,\dots,g$ the $j_{\text{th}}$ 
Chern class of~$\mathbb{E}_{g,n}$.
The following integrals
\be \label{Hodgeintegralsdef}
 \int_{\overline{\M}_{g,n}} \psi_1^{k_1}\cdots\psi_n^{k_n}\lambda_{j_1} \cdots \lambda_{j_m}  =: \langle \lambda_{j_1} \cdots \lambda_{j_m} \tau_{k_1}\cdots\tau_{k_n}\rangle_{g}
\ee
are some rational numbers, called Hodge integrals of a point.
Here $n,m\geq 0$, $k_1,\dots,k_n\geq 0$, $j_1,\dots,j_m\geq 1$. 
From the degree-dimension counting, these rational numbers 
vanish unless
\be
j_1+\dots+j_m+k_1+\dots +k_n=3g-3+n.
\ee
The case $m=0$ in~\eqref{Hodgeintegralsdef} gives the Gromov--Witten (GW) invariants of a point. 
For this case, the degree-dimension counting reads $k_1+\dots+k_n=3g-3+n$. 
So one could simply write $\langle \tau_{k_1}\cdots\tau_{k_n}\rangle_{g}$ 
as $\langle \tau_{k_1}\cdots\tau_{k_n}\rangle$, which are the numbers used in~\eqref{M-g}.

It is appropriate to collect Hodge integrals into generating series. The genus~$g$ Hodge 
potential associated to $\lambda_{i_1}\cdots\lambda_{i_m}$ is defined 
as the following generating series of Hodge integrals:
\be
\HH_g(\lambda_{i_1}\cdots\lambda_{i_m};\bt):=\sum_{n=0}^\infty \frac{1}{n!} \sum_{k_1,...,k_n\geq 0} 
\bigl\langle \lambda_{i_1} \cdots \lambda_{i_m}\tau_{k_1}\cdots\tau_{k_n}\bigr\rangle_g \, t_{k_1}\cdots t_{k_n},
\ee
where $\bt=(t_0,t_1,t_2,\dots)$.
Denote by ${\rm ch}_r:={\rm ch}_r(\mathbb{E}_{g,n})$, $r\geq 0$ components of the Chern character of~$\mathbb{E}_{g,n}$. 
We call the generating series 
\be
\HH_g(\bt;\bs):=\sum_{m,n=0}^\infty \frac{1}{m! \, n!} \sum_{j_1,...,j_m\geq 1 \atop k_1,...,k_n\geq 0} \langle {\rm ch}_{2j_1-1}\cdots {\rm ch}_{2j_m-1} \tau_{k_1}\cdots\tau_{k_n}\rangle_{g} \, s_{j_1}\cdots s_{j_m}\, t_{k_1}\cdots t_{k_n}
\ee
the genus~$g$ Hodge potential. Here $\bs=(s_1,s_2,\dots)$.
The restriction $\HH_g(\bt;{\bf 0})=:\F_g(\bt)$ is called the genus~$g$ GW potential. 
We also define the Hodge potential $\HH$ and the GW potential~$\F$ by
 \[ \HH=\HH(\bt;\bs;\e):=\sum_{g\geq 0} \e^{2g-2}\HH_g(\bt;\e), \quad \F=\F(\bt;\bs;\e):=\sum_{g\geq 0} \e^{2g-2} \F_g(\bt;\e).\]
Their exponentials 
\[ Z_E=Z_E(\bt;\bs;\e):=\exp\left(\HH(\bt;\bs;\e)\right), \quad Z=Z(\bt; \e):=\exp \left(\F(\bt;\e)\right)\]
are called the partition functions of Hodge integrals and of GW invariants, respectively. 

It was conjectured by Witten~\cite{Witten} and proved by Kontsevich~\cite{Kontsevich} that 
$Z$ is a particular tau-function of the KdV hierarchy (the Witten--Kontsevich (WK) theorem). 
$Z$ is now  
also known as the WK tau-function.
Define $D_k$ as the following linear differential operators~\cite{FP}:
\be
D_k:=\sum_{p\geq 0}  t_p \frac{\p }{\p t_{p+2k-1}} - \frac{\e^2}{2} \sum_{j=0}^{2k-2} (-1)^j \frac{\p^2 }{\p t_j \p t_{2k-2-j}},\quad k\geq 1.
\ee
Faber--Pandharipande~\cite{FP} proved that the partition function of Hodge integrals 
$Z_E(\bt;\bs;\e)$ is the unique power series solution to the following linear equations
\be\label{thefaberpand1}
\frac{\p Z_E}{\p s_k}=-\frac{B_{2k}}{(2k)!} D_k \bigl( Z_E \bigr),\qquad k\geq 1
\ee
along with the initial condition
\be\label{thefaberpand2}
Z_E(\bt;\bdzero;\e)=Z(\bt;\e).
\ee
This unique solution has the form
\be\label{G-exp}
Z_E(\bt;\bs;\e)=\exp\biggl(-\sum_{k\geq 1} \frac{B_{2k}}{(2k)!}s_k D_k \biggr) \bigl(Z(\bt;\e)\bigr).
\ee

\begin{lem} \label{string-dilaton} The power series $Z_E$ satisfies the following two linear equations:
\begin{align}
& {\rm (string~equation)}\quad \sum_{p\geq 0} t_{p+1} \frac{\p Z_E}{\p t_{p}} + \frac{ t_0^2 }{2\e^2} Z_E+\frac{s_1}{24} Z_E=\frac{\p Z_E}{\p t_0},\label{string-hodge}\\
& {\rm (dilaton~equation)}\quad \sum_{p\geq 0} t_p \frac{\p Z_E}{\p t_p}+\e \frac{\p Z_E}{\p \e}+\frac{1}{24}Z_E=\frac{\p Z_E}{\p t_0}.\label{dilaton}
\end{align}
\end{lem}
\begin{prf} We have
\be
\langle\gamma \tau_{k_1}\cdots \tau_{k_n} \tau_0 \rangle_g
=\sum_{j=1}^n \, \left\langle \gamma \, \tau_{k_1-\delta_{1j}} \cdots \tau_{k_n-\delta_{nj}} \right\rangle_g,
\ee
where $\gamma=\lambda_{i_1}\cdots \lambda_{i_\ell},\,i_1,\dots, i_\ell\geq 1$ ($\gamma:=1$ if $\ell=0$), and $\langle\gamma\, \tau_{j_1}\cdots \tau_{j_n}\rangle_g:=0$ if $\{j_1,\dots,j_n\}$ contain negative integers.
It should be noted that there are two exceptional cases:
\begin{itemize}
\item[(a)] $g=0,\,n=2, \,  \gamma=1$. We have $\langle\tau_0^3\rangle_0=1$.
\item[(b)] $g=1, \, n=0, \, \gamma=s \, \lambda_1$. We have 
\be \label{well}
\langle \lambda_1  \tau_0\rangle_1=\frac{s}{24}.
\ee
\end{itemize}
They correspond to the terms 
$\frac{t_0^2}{2\e^2}Z_\mathbb{E}$, $\frac{s_1}{24} Z_\mathbb{E}$ in~\eqref{string-hodge}, respectively. 
This proves the string equation. Similarly, one can show that
\be\label{dilaton2}
\langle \gamma  \tau_{k_1}\cdots \tau_{k_n} \tau_1\rangle_g=(2g-2+n) \langle \gamma  \tau_{k_1}\cdots \tau_{k_n}\rangle_g.
\ee
There is one exceptional case: $g=n=1$, $\gamma=1$.  We have
$\langle\tau_1\rangle_1=\frac{1}{24}$. This proves~\eqref{dilaton}.
\end{prf}

We call $\langle\gamma \tau_{k_1}\cdots \tau_{k_n}\rangle_g$ 
a {\it primitive} Hodge integral of a point, if $k_1,\dots,k_n\geq 2$. 

In~\cite{DZ-norm} Dubrovin--Zhang (DZ) introduced the quasi-triviality approach to construct {\it the} integrable hierarchy 
for Gromov--Witten invariants of an arbitrary smooth projective 
variety~$X$ with semisimple quantum cohomology. For the case $X={\rm a~point}$, 
the DZ hierarchy coincides with the KdV hierarchy. The interesting fact is that 
{\it one particular solution can contain the full information of the unique integrable system \cite{DZ-norm, Du2, BPS, DLYZ, DLYZ-2}}. 
Recently, the integrable hierarchy for Hodge integrals called the Hodge hierarchy 
({\it aka} the DZ hierarchy for Hodge integrals) was constructed in~\cite{DLYZ} (see also~\cite{BPS}) 
by using the quasi-triviality approach. 
Let us review the construction. 
The genus~0 and genus~1 Hodge potentials have the form
\begin{align}
& \HH_0(\bt;\bs)=\F_0(\bt)=\sum_{n\geq 3} \frac1{n(n-1)(n-2)}\sum_{k_1+\cdots+ k_n=n-3} \frac{t_{k_1}}{k_1!}\cdots \frac{t_{k_n}}{k_n!}, \label{H0} \\
& \HH_1(\bt;\bs)= H_1\bigl(v^{\rm top}(\bt),v^{\rm top}_x(\bt)\bigr), \label{H1}
\end{align}
where $H_1(v,v_x):=\frac1{24} \log v_x + \frac1{24} v$. 
For higher genera, the following lemma is useful. 
\begin{lem} [Theorem 1.3 of \cite{DLYZ}] \label{thm-dlyz} 
For each $g\geq 2$, there exist a unique element 
$$H_g=H_g(v,v_1,\dots,v_{3g-2};s_1,\dots,s_g)$$ 
in $C^\infty(v)\bigl[v_1,v_1^{-1};v_2,\dots,v_{3g-2};s_1,\dots,s_g\bigr]$ satisfying 
\begin{align}
& \HH_g(\bt;\bs)=H_g\bigl(v(\bt), v_x(\bt), v_2(\bt), \dots, v_{3g-2}(\bt); s_1, \dots, s_g\bigr),  \label{Hg-dlyz1}\\
& \deg H_g=2g-2,\label{Hg-dlyz2}
\end{align}
where $v(\bt):=v^{\rm top}(\bt)$.
Moreover, define $\overline{\deg} \, v_k:=k-1$ and $\overline{\deg} \, s_k:=2k-1$, $k\geq 1$, then
\beq\label{deghg}
\overline{\deg} \, H_g \leq 3g-3. 
\eeq
\end{lem}
\noindent A straightforward proof of this lemma by using the string and dilaton equations (cf. Lemma~\ref{string-dilaton}) 
is given in Appendix~\ref{app1014}.

By Lemma~\ref{thm-dlyz} we know that the change of the dependent variable 
\be\label{quasimiurahodge}
v\mapsto w=v+\sum_{g\geq 1} \e^{2g} \p_x^2 \bigl(H_g\bigr) 
\ee
is a quasi-Miura transformation. Here $\p_x=\sum_{k\geq 0} v_{k+1} \p_{v_k}$.
The Hodge hierarchy is defined as the 
PDE system obtained from the Riemann hierarchy~\eqref{riemann} 
under the quasi-Miura transformation~\eqref{quasimiurahodge}.
Each member of the Hodge hierarchy is proven to have the form~\eqref{normal-form}, 
and so is integrable \cite{DLYZ,BPS}. 
We note that the uniqueness part of~Lemma~\ref{thm-dlyz} was implicit in 
Theorem~1.3 of~\cite{DLYZ}. (It can be deduced from the statement in Theorem~1.3 of~\cite{DLYZ} that 
$H_g$ is independent from the choice of a solution; or, it can be directly proved by using 
an argument similar to that appears in the proof of Theorem~4.2.4 in the first 
arXiv preprint version of~\cite{DLYZ-2}, which uses the ``transcendental property" 
of~$v^{\rm top}(\bt)$.)

\medskip

\noindent \textbf{Theorem (\cite{DLYZ}).} {\it $Z_E$
 is a particular tau-function of the Hodge hierarchy.}

\begin{lem} \label{ind-v-lem}
The following formulae hold true:
\be\label{ind-v}
\frac{\p H_g}{\p v}=0, \quad \forall\, g\geq 2.
\ee
\end{lem}
\begin{prf}
Take $v=v(\bt)=v^{\rm top}(\bt)$.
Equation~\eqref{string-hodge} implies that for $g\geq 2$,
$\sum_{p\geq 0} t_{p+1} \frac{\p \mathcal{H}_g}{\p t_{p}}=\frac{\p \mathcal{H}_g}{\p{t_0}}$.
Substituting~\eqref{Hg-dlyz1}  in this equation and using the Riemann hierarchy we obtain
\be
\sum_{k=0}^{3g-2} \frac{\p H_g}{\p v_k} \p_x^{k+1} \Biggl(\sum_{p\geq 0} t_{p+1}\frac{v^{p+1}}{(p+1)!}\Biggr)=\sum_{k=0}^{3g-2} \frac{\p H_g}{\p v_k}\p_x^{k+1}(v).
\ee
Substituting equation~\eqref{el} with $c_p=\delta_{p,1}$ in the above equation we arrive at~\eqref{ind-v}.
\end{prf}

\smallskip

We now formulate a theorem providing more accurate expressions for $H_g$, $g\geq 2$.
\begin{thm}\label{quasi-Hodge} The genus $g$ Hodge potentials have the following expressions:
\eqa
&&\HH_0(\bt;\bs)=\F_0(\bt),\label{HH0}\\
&&\HH_1(\bt;\bs)=M_1(v_x(\bt))+\frac{s_1}{24} v(\bt),\label{HH1}\\
&&\HH_g(\bt;\bs)=M_g(v_x(\bt),\dots,v_{3g-2}(\bt))+\sum_{m=1}^\infty \frac{1}{m!} \sum_{1\leq j_1,\dots,j_m\leq  g} s_{j_1}\cdots s_{j_m} v_x(\bt)^{-g+1+2\sum_{a=1}^m j_a -m}\nn\\
&&\qquad \qquad \sum_{\lambda,\mu\in\mathbb{Y}_{3g-3+m-2\sum_{a=1}^m j_a}} \frac{\langle {\rm ch}_{2j_1-1}\cdots {\rm ch}_{2j_m-1}\, \tau_{\lambda+1}\rangle_g}{m(\lambda)!} \,  Q^{\lambda\mu} \, \frac{v_{\mu+1}(\bt)}{v_x(\bt)^{l(\mu)}},\quad g\geq 2.\label{HHg}
\eeqa
Here, $M_g$, $g\geq 1$ are defined in \eqref{M1-kdv}--\eqref{M-g}, and
$
v(\bt)=\sum_{k\geq 1} \frac{1}{k}\sum_{p_1+\cdots+p_k=k-1} \frac{t_{p_1}}{p_1!}\cdots\frac{t_{p_k}}{p_k!}
$.
\end{thm}
%Formula~\eqref{HH0} is an easy fact from topology, and formula~\eqref{HH1} has been obtained in~\cite{DLYZ}.
\begin{prf} The $g=0,1$ cases are already known (cf. \eqref{H0} and~\eqref{H1}). %in Lemma~\ref{thm-dlyz}.
For $g\geq 2$, we have 
\eqa
\HH_g(\gamma; \bt)&=&\sum_{n=0}^\infty \frac{1}{n!}\sum_{k_1,\dots,k_n\geq 0 \atop k_1+\cdots +k_n+\deg \gamma= 3g-3+n} \langle\gamma\, \tau_{k_1}\cdots \tau_{k_n}\rangle_{g,n}\, t_{k_1} \cdots t_{k_n}. \label{hg}
\eeqa
Here, $\gamma:={\rm ch}_{2i_1-1}\cdots {\rm ch}_{2i_m-1},\,i_1,\dots,i_m\geq 1$ ($\gamma:=1$, if $m=0$). By definition we know that
\be
\HH_g(\bt;\bs)=\sum_{m=0}^\infty \frac{1}{m!} \sum_{1\leq i_1,\dots, i_m\leq g} 
\HH_g({\rm ch}_{2i_1-1}\cdots {\rm ch}_{2i_m-1};\bt)\, s_{i_1}\cdots s_{i_m}.
\ee
According to Lemma~\ref{thm-dlyz} there exist functions $H_g(\gamma;v_x,\dots,v_{3g-2})$ such that
\be
\HH_g(\gamma;\bt)=H_g\bigl(\gamma;v_x(\bt),\dots, v_{3g-2}(\bt)\bigr),\quad g\geq 2,
\ee
where $H_g(\gamma;v_x,\dots,v_{3g-2})\in C^\infty(v)\bigl[v_1,v_1^{-1};v_2,\dots,v_{3g-2}\bigr]$.
Taking $t_0=0$ in equation~\eqref{hg} we find
\eqa
\HH_g(\gamma; 0,t_1,t_2,\dots)&=&\sum_{n=0}^\infty \frac{1}{n!}\sum_{k_1,\dots,k_n\geq 1 \atop k_1+\cdots +k_n+\deg \gamma= 3g-3+n} 
\langle\gamma\, \tau_{k_1}\cdots \tau_{k_n}\rangle_{g,n} \, t_{k_1} \cdots t_{k_n}\nn\\
%&=&\sum_{n=0}^\infty \frac{1}{n!} \sum_{\ell(\lambda)=n\atop \lambda\in\mathbb{Y}_{3g-3+n-\deg\gamma}} 
%\binom{n}{m_1(\lambda),m_2(\lambda),\dots} \, \langle\gamma\, \tau_\lambda\rangle_{g,n}\, t_{\lambda}\nn\\
&=&\sum_{m_1,m_2,m_3,\dots \geq 0 \atop \sum (i-1)m_i =3g-3-\deg \gamma} \bigl\langle\gamma\, \tau_1^{m_1} \tau_2^{m_2} \cdots \bigr\rangle_{g}\, \prod_{i=1}^\infty \frac{t_i^{m_i}}{m_i!}.\label{abc}
\eeqa
Substituting the dilaton equation \eqref{dilaton} into \eqref{abc} we obtain
\eqa
\HH_g(\gamma; 0,t_1,t_2,\dots)
&=&\sum_{m_1,m_2,m_3,\dots \geq 0 \atop \sum (i-1)m_i =3g-3-\deg \gamma} \frac{(\sum m_i+2g-3)!}{(\sum m_i+2g-3-m_1)!} \, \langle\gamma\, \tau_2^{m_2} \tau_3^{m_3} \cdots \rangle_{g}\, \prod_{i=1}^\infty \frac{t_i^{m_i}}{m_i!}\nn\\
&=&\sum_{m_2,m_3,m_4,\dots \geq 0 \atop \sum_{i=2}^\infty  (i-1)m_i =3g-3-\deg \gamma} \frac{\langle\gamma\, \tau_2^{m_2}\tau_3^{m_3} \cdots \rangle_{g}}{(1-t_1)^{\sum_{i=2}^\infty m_i+2g-2}}\, \prod_{i=2}^\infty \frac{t_i^{m_i}}{m_i!}\nn\\
&=&\sum_{\lambda\in\mathbb{Y}_{3g-3-\deg\gamma}} \frac{\langle\gamma\, \tau_{\lambda+1} \rangle_{g}}{(1-t_1)^{\ell(\lambda)+2g-2}}\, \frac{t_{\lambda+1}}{m(\lambda)!}.\label{def}
\eeqa
Here we have used Newton's binomial identity
$(1-x)^{-1-k}=\sum_{s=0}^\infty \binom{s+k}{k} x^s$.
Substituting formula~\eqref{tv} and~\eqref{t1} into~\eqref{def} we obtain
\eqa
\HH_g(\gamma; 0, t_1,t_2,\dots)&=&\sum_{\gamma\in\mathbb{Y}_{3g-3-\deg\gamma}} 
\langle \gamma\, \tau_{\lambda+1}\rangle_g \, (v_x^s)^{2g-2+\ell(\gamma)}\, \frac{(-1)^{\ell(\lambda)}}{m(\lambda)!} \,  \sum_{\mu\in\mathbb{Y}_{|\lambda|}} \, Q^{\lambda\mu}\,\frac{v^s_{\mu+1}}{(v_x^s)^{l(\mu)+|\lambda+1|}} \nn\\
&=&(v_x^s)^{-g+1+\deg\gamma}\sum_{\lambda,\mu\in\mathbb{Y}_{3g-3-\deg\gamma}} \frac{\langle \gamma\, 
\tau_{\lambda+1}\rangle_g}{m(\lambda)!} \, Q^{\lambda\mu}\, \frac{v^s_{\mu+1}}{(v_x^s)^{l(\mu)}}. \label{e1}
\eeqa
Finally, due to Lemma \ref{thm-dlyz} and Lemma \ref{ind-v-lem}, $\mathcal{H}_g(\gamma;\bt)$ must have the form
\be
\mathcal{H}_g(\gamma;\bt)=\sum_{q\geq 0} v_x^{-g+1+\deg\gamma+q}\sum_{\mu\in\mathbb{Y}_{3g-3-\deg\gamma-q}} c^{g,q}_{\mu} \, \frac{v_{\mu+1}}{v_x^{l(\mu)}}.
\ee
Taking $t_0=0$ we have
\be
\mathcal{H}_g(\gamma;0,t_1,t_2,\dots)=\sum_{q\geq 0} (v_x^s)^{-g+1+\deg\gamma+q}\sum_{\mu\in\mathbb{Y}_{3g-3-\deg\gamma-q}} c^{g,q}_{\mu} \, \frac{v^s_{\mu+1}}{(v_x^s)^{l(\mu)}}.\label{e2}
\ee
Comparing equations \eqref{e1} and \eqref{e2} we find
\be
c_{\mu}^{g,q}=0,\,\mbox{if }q\geq 1;\qquad c_{\mu}^{g,q}=\sum_{\lambda\in\mathbb{Y}_{3g-3-\deg\gamma}} \frac{\langle \gamma\, \tau_{\lambda+1}\rangle_g}{m(\lambda)!} \, Q^{\lambda\mu},\,\mbox{if } q=0.
\ee
The theorem is proved.
\end{prf}

\begin{cor} \label{degree-homo} 
For $g\geq 2$,
$H_g(v_x,\dots,v_{3g-2};s_1,\dots,s_g)$ is homogenous of degree~$3g-3$ with respect to~$\overline{\deg}$.
\end{cor}

It follows from Mumford's relation
\be
\Lambda_g^\vee(s) \Lambda_g^\vee(-s) = (-1)^g s^{2g}
\ee
as well as from the relationship between Schur basis and power sum basis of symmetric functions  
that the infinite set $\{\lambda_{j_1}\cdots\lambda_{j_n}\}$ and the infinite set $\{{\rm ch}_{2i_1-1}\cdots {\rm ch}_{2i_m-1}\}$ span the same infinite dimensional vector space. Here $\Lambda_g^\vee(s) :=\sum_{i=0}^g (-s)^i \lambda_{g-i},\,\lambda_0:=1$.
Therefore, for any linear combination 
$\gamma=\sum_n \sum_{j_1,\dots,j_n} A_{j_1,\dots,j_n} \lambda_{j_1}\cdots\lambda_{j_n}= 
\sum_m \sum_{i_1,\dots,i_m}  B_{i_1,\dots,i_m} {\rm ch}_{2i_1-1}\cdots {\rm ch}_{2i_m-1},$ 
the function $H_g(\gamma;v,v_x,\dots,v_{3g-2};s_1,\dots,s_g)$ is also defined (via linear combination).

\begin{emp}
$\gamma=\lambda_g\lambda_{g-1}\lambda_{g-2}$. Noting that 
\be
\langle \lambda_g\lambda_{g-1}\lambda_{g-2}\rangle_g=\frac{1}{2(2g-2)!}\frac{|B_{2g-2}|}{2g-2}\frac{|B_{2g}|}{2g}, \nn
\ee
we have
\be
H_g(\lambda_g\lambda_{g-1}\lambda_{g-2};v_1,\dots,v_{3g-2})=\frac{1}{2(2g-2)!}\frac{|B_{2g-2}|}{2g-2}\frac{|B_{2g}|}{2g}v_x^{2g-2},\qquad g\geq 2. \nn
\ee
\end{emp}

\begin{emp}
$\gamma=\lambda_g$. The $\lambda_g$-conjecture proven for example in~\cite{FP} tells that
\be
\langle\lambda_g \,\tau_{k_1}\cdots \tau_{k_n}\rangle_g=\frac{2^{2g-1}-1}{2^{2g-1}}\frac{|B_{2g}|}{(2g)!}\frac{(2g-3+n)!}{k_1! \cdots k_n!}. \nn
\ee
Therefore, 
\eqa
H_g(\gamma;v_1,\dots,v_{3g-2})&=&\frac{2^{2g-1}-1}{2^{2g-1}}\frac{|B_{2g}|}{(2g)!} \, v_x \sum_{\lambda,\mu\in\mathbb{Y}_{2g-3}}\frac{(2g-3+\ell(\lambda))!}{z_{\lambda+1}}\frac{v_{\mu+1}}{v_x^{\ell(\mu)}} Q^{\lambda\mu}\nn\\
&=&\frac{2^{2g-1}-1}{2^{2g-1}}\frac{|B_{2g}|}{(2g)!} \sum_{\lambda,\mu\in\mathbb{Y}_{2g-3}}(-1)^{\ell(\lambda)} L(\lambda) \, Q^{\lambda\mu}\,\frac{v_{\mu+1}}{v_x^{\ell(\mu)-1}}. \nn
\eeqa
Noting that due to e) and f) of Lemma~\ref{QQ-pro}, $Q^{\lambda\mu}$ satisfy the following property:
\be
\sum_{\lambda\in\mathbb{Y}_{2g-3}}(-1)^{\ell(\lambda)} L(\lambda) \, Q^{\lambda\mu}=\delta_{\mu,(2g-3)},\quad g\geq 2. \nn
\ee
Hence we obtain
\be
H_g\bigl(\lambda_g;v_1,\dots,v_{3g-2}\bigr)=\frac{2^{2g-1}-1}{2^{2g-1}}\frac{|B_{2g}|}{(2g)!} \, v_{2g-2},\qquad g\geq 2. \nn
\ee
\end{emp}

\begin{emp}
$\gamma=\lambda_g\lambda_{g-1}$. Getzler--Pandharipande~\cite{GP} proved that for $k_1,\dots,k_n\geq 1$,
\be
\langle\lambda_g \lambda_{g-1}\,\tau_{k_1}\cdots \tau_{k_n}\rangle_g=\frac{(2g+n-3)!}{(2k_1-1)!!\cdots(2k_n-1)!!}\frac{|B_{2g}|}{2^{2g-1}(2g)!}. \nn
\ee
Therefore we have
\be
H_g(\lambda_g\lambda_{g-1};v_1,\dots,v_{3g-2})=\frac{|B_{2g}|}{2^{2g-1}(2g)!} \, v_x^g \sum_{\lambda,\mu\in\mathbb{Y}_{g-2}}\frac{(2g-3+\ell(\lambda))!}{\prod_{i\geq 1} (2i+1)!!^{m_i(\lambda)} \, m(\lambda)!}\,Q^{\lambda\mu}\, \frac{v_{\mu+1}}{v_x^{\ell(\mu)}}.\nn
\ee
\end{emp}

\section{Solutions to Problems A,B,C}\label{proof}
In this section we provide solutions to Problems A,B,C using the Witten--Kontsevich theorem, 
Buryak's theorem~\cite{Buryak}, and results of~\cite{DLYZ-2}, respectively.

\medskip

\noindent {\bf A solution to Problem A -- Proof of Theorem~\ref{quasi-KdV}.} ~ 
Using Theorem~\ref{quasi-Hodge} it is easy to verify that $H_1(v_x; \bdzero)=M_1(v_x) = \frac1{24} \log v_x$, and  
\be
H_g(v_x,\dots,v_{3g-2}; \bdzero)=M_g(v_x,v_{xx},\dots,v_{3g-2}),  
\quad g\geq 2. \nn
\ee 
By using the Witten--Kontsevich theorem and Lemma~\ref{thm-dlyz}, we know that
\be\label{coor-v-uA}
 u^{\rm top}(\bt;\e):=
 v^{\rm top}(\bt)+\sum_{g=1}^\infty \e^{2g} \p_x^2 H_g\bigl(v^{\rm top}_x(\bt),\dots,v^{\rm top}_{3g-2}(\bt); {\bf 0}\bigr)
\ee
satisfies the KdV hierarchy. In particular it satisfies the KdV equation~\eqref{KdV-eqn}. 
We then deduce from the transcendental property of~$v^{\rm top}(\bt)$ that for any solution $v(\bt)$ to the Riemann hierarchy in 
$\CC[[\bt]]$ satisfying $v_x(\bt) \neq 0$, the function $u(\bt;\e)$ defined by~\eqref{coor-v-uA} with $v^{\rm top}$ being replaced by~$v$
also satisfies the KdV hierarchy. Theorem~\ref{quasi-KdV} is proved. $\hfill\square$

For  $\ell(\lambda)=1$, it is well-known (see for example~\cite{F}) that 
\be \langle\tau_{3g-2}\rangle=\frac{1}{24^g\, g!},\quad g\geq 1.\ee
For $\ell(\lambda)\geq 2$, a recently formula \cite{BDY} gives
\be \langle\tau_{\lambda+1}\rangle=(-1)^{\ell(\lambda)+1} \prod_{i=1}^{\ell(\lambda)} \res{z_{i}=\infty} d z_i \, z_i^{2\lambda_i+3} \left(\sum_{r\in S_{\ell(\lambda)}}  \frac{\Tr \bigl(R(z_{r_1})\cdots 
R(z_{r_{\ell(\lambda)}})\bigr)}{\ell(\lambda) \, \prod_{j=1}^{\ell(\lambda)}(z_{r_j}^2-z_{r_{j+1}}^2 )} + \delta_{\ell(\lambda),2}\frac{z_1^2+z_2^2}{(z_1^2-z_2^2)^2}\right). \nn
\ee
Here for a permutation $r=[r_1,\dots,r_\ell]$ in $S_\ell$, $r_{\ell+1}:=r_1$, and  
\be
R(z)=\frac{1}{2}\left(
\begin{array}{cc}
-\sum_{g=1}^\infty \frac{(6g-5)!!}{24^{g-1}\, (g-1)!} z^{-6g+4} & -2 \sum_{g=0}^\infty \frac{(6g-1)!!}{24^g\, g!} z^{-6g}\\
\\
2 \sum_{g=0}^\infty\frac{6g+1}{6g-1} \frac{(6g-1)!!}{24^g\, g!} z^{-6g+2} &  \sum_{g=1}^\infty \frac{(6g-5)!!}{24^{g-1}\, (g-1)!} 
z^{-6g+4}\\
\end{array}
\right).
\ee

\medskip

\noindent {\bf A solution to Problem B}. ~~
Let  $\Lambda_g(s) :=\sum_{i=0}^g s^i \lambda_{i}$ be the Chern polynomial of the Hodge bundle. 
By using Buryak's theorem~\cite{Buryak} and Lemma~\ref{thm-dlyz} we know that 
\be
 w(\bt;\e):=\sum_{g=0}^\infty \e^{2g} 
 \p_x^2 \Bigl(\HH_g\bigl(\Lambda(s);\bt\bigr)\Bigr)
 =v^{\rm top}(\bt)+\sum_{g=1}^\infty \e^{2g} \p_x^2 \Bigl(H_g\bigl(\Lambda(s); v^{\rm top}(\bt), v^{\rm top}_x(\bt),\dots,v^{\rm top}_{3g-2}(\bt)\bigr)\Bigr) 
 \label{coor-v-wB}
\ee
is a particular solution to an explicit deformation of the intermediate long wave (ILW) hierarchy. 
We deduce from the transcendental property of~$v^{\rm top}(\bt)$ that for any solution $v(\bt)$ to the Riemann hierarchy in 
$\CC[[\bt]]$ satisfying $v_x(\bt) \neq 0$, the function~$w(\bt;\e)$ defined by~\eqref{coor-v-wB} 
with $v^{\rm top}$ being replaced by~$v$ is a solution to the explicit deformation of the ILW hierarchy. 
In other words, the composition of the quasi-Miura transformation
\begin{align}
& w=v+\sum_{g=1}^\infty \e^{2g} \p_x^2 H_g(\Lambda(s); v_x,\dots,v_{3g-2}), \label{coor-v-w} \\
& H_1(\Lambda(s);v,v_x)=\frac{1}{24}\log v_x+\frac{s}{24} v,\\
& H_g(\Lambda(s);v_1,\dots,v_{3g-2})=\sum_{k=0}^g s^g \, v_x^{-g+1+k}\sum_{\lambda,\mu\in\mathbb{Y}_{3g-3-k}} \frac{\langle \lambda_g 
\tau_{\lambda+1}\rangle_g}{m(\lambda)!} \, Q^{\lambda\mu}\, \frac{v_{\mu+1}}{v_x^{\ell(\mu)}},\quad g\geq 2 \label{derivationhg}
\end{align}
with the Miura type transformation
\be\label{cooruwmiura}
u=w+ \sum_{g\geq 1} \frac{(-1)^g}{2^{2g}(2g+1)!}\e^{2g}s^{g} w_{2g}
\ee
gives the quasi-triviality of the ILW equation (equivalently of the whole ILW hierarchy)
\be
u_t=u u_x+ \sum_{g\ge1}\e^{2g}s^{g-1}\frac{|B_{2g}|}{(2g)!}u_{2g+1}.
\ee
Here $t=t_1$. In the derivation of~\eqref{derivationhg} we used Theorem~\ref{quasi-Hodge} and the relationship
\be
H_g(v,v_x,\dots,v_{3g-2};\bs) \mid_{s_k=(2k-2)! s^{2k-1}, k\geq 1} 
= 
H_g(\Lambda(s); v(\bt), v_x(\bt),v_2(\bt),\dots),\quad g\geq 1.
\ee
We conclude that the above~\eqref{coor-v-wB}--\eqref{cooruwmiura} give a solution to Problem~B in terms of 
composition of an explicit quasi-Miura transformation with an explicit Miura type transformation.

\medskip

\noindent {\bf A solution to Problem C.} ~~
It was conjectured in~\cite{DLYZ} that the quasi-Miura map 
\be\label{coor-v-w}
w=v+\sum_{g=1}^\infty \e^{2g} \p_x^2 H_g \bigl(\Lambda(s)\Lambda(-2s)^2; v_x,v_{xx},\dots\bigr)
\ee
with $s=1$ gives rise to an explicit deformation of the discrete KdV hierarchy. This conjecture was 
proven in~\cite{DLYZ-2}. It says, more precisely, that for $s=1$ the composition of the following three transformations
\begin{align}
& w=v+\sum_{g=1}^\infty \e^{2g} \p_x^2 H_g(\Lambda(s)\Lambda(-2s)^2; v_x,v_{xx},\dots), \label{coor-v-w} \\
& H_1\bigl(\Lambda(s)\Lambda(-2s)^2;v_x\bigr)=\frac{1}{24}\log v_x-\frac{s}{8} v,  \nn\\
& H_g\bigl(\Lambda(s)\Lambda(-2s)^2;v_1,\dots,v_{3g-2}\bigr)
=\sum_{k=0}^{3g-3} s^{k}\, v_x^{-g+1+k}   \sum_{k_1+k_2+k_3=k\atop 0\leq k_1,k_2,k_3\leq g} \nn\\
&\qquad\qquad\qquad\qquad (-2)^{k_2+k_3}\sum_{\rho,\mu\in\mathbb{Y}_{3g-3-k}} \frac{\langle \lambda_{k_1}\lambda_{k_2}\lambda_{k_3}
\tau_{\rho+1}\rangle_g}{m(\rho)!} \, Q^{\rho\mu}\, \frac{v_{\mu+1}}{v_x^{l(\mu)}},\qquad g\geq 2, \label{derivationHg2}
\end{align}
\be
\tilde{w}=\frac{w}{2}, \qquad \mbox{and}  \qquad
u=\tilde{w}+\sum_{k=1}^\infty \e^{2k} \frac { 3^{2k+2}-1 } { (2k+2)!
4^{k+1} } \tilde{w}_{2k}
\ee
gives the quasi-triviality of the discrete KdV equation
\be
u_t=\frac{1}{\e}\Bigl(e^{u(x+\e)}-e^{u(x-\e)}\Bigr).
\ee
In the derivation of~\eqref{derivationHg2} we have used Theorem~\ref{quasi-Hodge} and the relationship
\be
H_g(v_x,\dots,v_{3g-2};\bs)|_{s_k=-(4^k-1)(2k-2)! s^{2k-1}}=H_g\bigl(\Lambda(s)\Lambda(-2s)^2;v_x,\dots,v_{3g-2}\bigr),\quad g\geq 1.
\ee

\section{Quasi-triviality of the Burgers hierarchy}\label{Burgers}
The Burgers hierarchy
\be\label{Burgers-hierarchy}
u_{t_n}=\frac{1}{(n+1)!} \p_x \circ (\e\,\p_x+u)^n (u),\quad n\geq 0
\ee
is an integrable deformation of the Riemann hierarchy, whose first member
coincides with the Burgers equation~\eqref{Burgers}. 
Here, as before we identify $t_0$ with~$x$. 
We call a function $\tau=\tau(\bt;\e)$ a {\it viscous tau-function} for the Burgers hierarchy if $\tau$ satisfies 
\be
\tau_{t_n}=\frac{\e^n}{(n+1)!} \, \tau^{(n+1)},\qquad n\geq 0.
\ee
For $\tau(\bt;\e)$ being a viscous tau-function for the Burgers hierarchy, one can check that the function
$u=u(\bt;\e):=\e\, \p_x \log \tau(\bt;\e)$ 
satisfies~\eqref{Burgers-hierarchy}. So we also call $\tau$ the tau-function of~$u$.
On the other hand, for any fixed solution~$u=u(\bt;\e)\in \mathbb{C}[[\bt;\e]]$ to the Burgers hierarchy, 
 the tau-function $\tau\in\mathbb{C}((\e))[[\bt]]$ of~$u$ exists, and is unique up to 
 multiplying by a non-zero constant (which can depend on~$\e$).

The partition function $Z^{1D}(\bt;\e)$ of 1D gravity (toy model of quantum field theory) 
\be
Z^{1D}(\bt;\e):=\frac{1}{\sqrt{2\pi\e}} \int_{\mathbb{R}} 
e^{\frac{1}{\e} \bigl(-\frac{s^2}{2} + \sum_{n=0}^\infty \frac{t_n}{(n+1)!} s^{n+1}\bigr)}ds
\ee
is known to be a particular viscous tau-function\footnote{$Z^{1D}(\bt;\e)$ is also a tau-function of the KP hierarchy, 
where the KP times $T_1=t_0$, $T_2=t_1$, $T_3=t_2$, \dots.} of the Burgers hierarchy.  
The logarithm of $Z^{1D}(\bt;\e)$ admits the expansion 
\be
\log Z^{1D}(\bt;\e)=:\sum_{g=0}^\infty \e^{g-1} \mathcal{F}^{1D}_g(\bt).
\ee
By a direct computation of the Gaussian-type integral one obtains
\be
Z^{1D}(x,0,0,\cdots;\e)=e^{\frac{x^2}{2\e}}.
\ee
It follows that the initial value of the solution~$u^{1D}$ corresponding to $Z^{1D}(\bt;\e)$ is given by
\be\label{poly-ini}
u^{1D}(x,0,0,\cdots;\e)=x.
\ee

The series $v^{1D}(\bt):=\p_x \mathcal{F}^{1D}_0(\bt)$ satisfies the inviscid Burgers hierarchy (coinciding with the 
Riemann hierarchy), whose initial value reads
\be
v^{1D}(x,0,0,\cdots)=x.
\ee 
Therefore $v^{1D}(\bt)=v^{\rm top}(\bt)$. 
For $g\geq 1$, the following expressions for $\F^{1D}_g$ are known~\cite{Zhou}:
\begin{align}
&\mathcal{F}_1^{1D}(\bt)\big|_{t_0=0}=\frac{1}{2}\log(1-t_1),\\
&\mathcal{F}_g^{1D}(\bt)\big|_{t_0=0}=\sum_{\Gamma\in\mathcal{G}^c_{g,{\rm val}\geq3}} 
\frac{t_{\lambda(\Gamma)+1}}{| {\rm Aut} (\Gamma)| \, (1-t_1)^{E(\Gamma)}},\quad g\geq 2.
\end{align}
Here the summation is taken over all $g$-loop connected graphs 
whose vertices all have valences $\geq 3$, 
and $\lambda(\Gamma):= \bigl({\rm val}({\rm vertex}_1)-2,\dots, {\rm val}({\rm vertex}_{V(\Gamma)})-2\bigr)$. 

Introduce
\begin{align}
& F_1^{1D}(v_x) := -\frac 12 \log v_x, \label{deff11d}\\
& F_g^{1D}(v_1,\dots,v_{2g-2}) := \sum_{\mu\in\mathbb{Y}_{2g-2}} 
\sum_{\Gamma\in\mathcal{G}^c_{g, {\rm val}\geq3}} \frac{Q^{\lambda(\Gamma)\mu}}{|{\rm Aut}(\Gamma)|} \,\frac{v_{\mu+1}}{v_1^{l(\mu)+g-1}},\quad g\geq 2. \label{deffg1d}
\end{align}
Clearly, we have for $g\geq 2$ that $\deg F_g^{1D}=g-1,~ \overline{\deg} \,F_g^{1D}=2g-2$.
Applying Lemma \ref{tv-lem} we obtain that 
\begin{align}
& \mathcal{F}_1^{1D}(\bt)\big|_{t_0=0}= F_1^{1D}\bigl(v^s_x\bigr), \label{po-fg1} \\
& \mathcal{F}_g^{1D}(\bt)\big|_{t_0=0}= F_g^{1D}\bigl(v^s_1,\dots,v^s_{2g-2}\bigr),\quad g\geq 2, \label{po-fg}
\end{align}
where $v^s_k=v_k(\bt)|_{t_0=0}$, and we have used the Euler's formula $V(\Gamma)-E(\Gamma)=1-g$.
Using (181)--(183) of~\cite{Zhou} one can derive from \eqref{po-fg1}--\eqref{po-fg} the following identities: 
\begin{align}
& \mathcal{F}_g^{1D}(\bt)=F_1^{1D}\bigl(v^{1D}_1(\bt)\bigr), \label{po-f1-zzz-geq} \\
& \mathcal{F}_g^{1D}(\bt)=F_g^{1D}\bigl(v^{1D}_1(\bt),\dots,v^{1D}_{2g-2}(\bt)\bigr),\quad g\geq 2. \label{po-fg-zzz-geq} 
\end{align}
Using the transcendental property of~$v^{1D}(\bt)$, we arrive at the following solution to Problem~D.
\begin{thm} Quasi-triviality of the Burgers hierarchy \eqref{Burgers-hierarchy} has the expression
\be
u=v+\sum_{g=1}^\infty \e^g \p_x F^{1D}_g(v_1,v_2,\dots)=v-\e\biggl(\frac{v_{xx}}{2v_x}\biggr) + \mathcal{O}(\e^2),
\ee
where $F^{1D}_g$ are defined explicitly in~\eqref{deff11d}--\eqref{deffg1d}.
\end{thm}

\section{Conclusion}\label{further-rmks}

Quasi-triviality of the Hodge hierarchy of a point and primitive 
Hodge integrals of a point are related via $Q$-matrices. 
The relation consists of two parts.

\medskip

\noindent Part a). {\em From Hodge integrals to quasi-triviality of the Hodge hierarchy.} I.e., one can use primitive Hodge integrals
 to represent quasi-triviality of the Hodge hierarchy:
\begin{align}
&w=v+\sum_{g\geq 1} \e^{2g} \p_x^2 H_g, \nn \\
&H_1(v,v_x;\bs)=\frac{1}{24} \log(v_x)+\frac{s_1}{24} v, \nn \\
&H_g(v_1,\dots,v_{3g-2};\bs)=\sum_{\phi\in\mathbb{Y}\atop 1\leq \phi_1,\dots,\phi_{\ell(\phi)}\leq g} \,  \frac{s_{\phi}}{m(\phi)!} v_x^{-g+1+2|\phi| - \ell(\phi)}\nn\\
&\qquad \qquad \qquad \qquad \qquad \qquad \sum_{\lambda,\mu\in\mathbb{Y}_{3g-3+\ell(\phi)-2|\phi|}} \frac{\langle {\rm ch}_{2\phi-1}\, \tau_{\lambda+1}\rangle_g}{m(\lambda)!} \,  Q^{\lambda\mu} \, \frac{v_{\mu+1}}{v_x^{l(\mu)}},\quad g\geq 2, \nn
\end{align}
where ${\rm ch}_{2\phi-1}:= {\rm ch}_{2\phi_1-1}\cdots {\rm ch}_{2\phi_{\ell(\phi)}-1};\, {\rm ch}_{2\phi-1}:=1$ if $\ell(\phi)=0$.

\medskip

\noindent Part b). {\em From quasi-triviality of the Hodge hierarchy to Hodge integrals.}  I.e., 
one can use quasi-triviality of the Hodge hierarchy to represent primitive Hodge integrals. Write for $g\geq 2$,
\be
H_g(v_1,\dots,v_{3g-2};\bs)=\sum_{\phi\in\mathbb{Y}\atop 1\leq \phi_1,\dots,\phi_{\ell(\phi)}\leq g} \,  \frac{s_{\phi}}{m(\phi)!} v_x^{-g+1+2|\phi| - \ell(\phi)} \sum_{\mu\in\mathbb{Y}_{3g-3+\ell(\phi)-2|\phi|}} c_g^\mu({\rm ch}_{2\phi-1}) \frac{v_{\mu+1}}{v_x^{l(\mu)}}.
\ee
Then we have $\forall\,\lambda,\mu,\phi\in\mathbb{Y}$,
\be
\bigl\langle {\rm ch}_{2\phi-1} \tau_{\lambda+1} \bigr\rangle_g = m(\lambda)! \, \sum_{\lambda\in\mathbb{Y}_{3g-3+\ell(\phi)-2|\phi|}} \, 
Q_{\lambda\mu}\, c_g^\mu({\rm ch}_{2\phi-1}),
\ee
where $g\geq 2$.
We could also express Part~a) as
\be
c_g^\mu({\rm ch}_{2\phi-1})=\sum_{\lambda\in\mathbb{Y}_{3g-3+\ell(\phi)-2|\phi|}} \, 
Q^{\lambda\mu} \bigl\langle {\rm ch}_{2\phi-1} \tau_{\lambda+1} \bigr\rangle_g.
\ee

\begin{appendix}

\section{Straightforward proof of Lemma~\ref{thm-dlyz}} \label{app1014}
In Theorem~\ref{quasi-Hodge} we express the genus~$g$ Hodge potential~$\mathcal{H}_g$ ($g\geq 1$)
in terms of $v_m$, $m\geq 0$ with coefficients given by intersection numbers.
Here $v=v(\bt)$ is the topological solution~\eqref{vtop} to the dispersionless KdV hierarchy, 
and $v_m=v_m(\bt):=\p_x^m(v(\bt))$, $x=t_0$. Recall that our proof in Section~\ref{Hodge} uses 
Lemma~\ref{thm-dlyz} on the existence of 
jet-variables representation of $\mathcal{H}_g$; this lemma was 
proved in~\cite{DLYZ} based on the known existence of jet-variables 
representation~\cite{EYY,DZ-norm,Zhou} of the genus~$g$ Gromov-Witten 
potential of {\it a point} as well as on the uniqueness of the 
Faber-Pandharipande equations~\eqref{thefaberpand1}--\eqref{thefaberpand2}.
In this appendix, to make the results of this paper self-contained, we
give a straightforward proof of Lemma~\ref{thm-dlyz}. 
Recall that~\cite{IZ,EYY,GJV,Zhou} Itzykson-Zuber's formal power series are defined by
\beq\label{IZvar}
I_0 = I_0(\bt) := v(\bt), \qquad I_k = I_k(\bt) := \sum_{n\geq 0} t_{n+k} \frac{I_0^n}{n!} ~(k\geq1).
\eeq
Observing that $I_k = t_k + \cdots$, we know that~\eqref{IZvar} gives an invertible map between 
the $\bt$-variables $t_0,t_1,t_2,\dots$ and the $I$-variables $I_0,I_1,I_2,\dots$. The inverse map is given explicitly by~\cite{Zhou}
\beq\label{gentI}
t_k = \sum_{n=0}^\infty \frac{(-1)^n I_0^n}{n!} \, I_{n+k}.
\eeq
The following formula, which generalizes~Lemma~\ref{Zhou-lemma}, was 
derived in~\cite{Zhou}:
\be\label{generalizediv}
I_0=v, \quad I_k = \delta_{k,1} - \sum_{\mu\in \mathbb{Y}_{k-1}}  L(\mu) \, \frac{v_{\mu+1}}{v_1^{1+|\mu+1|}} ~(k\geq 1).
\ee
Formula~\eqref{generalizediv}  
can also be obtained by the (generalized) Lagrange inversion (cf. e.g.~\cite{L,J}). 
Combining~\eqref{generalizediv} with~\eqref{gentI} gives
\begin{align}
%& t_0 = -  \sum_{n=1}^\infty \frac{(-1)^n v^n}{n!}\sum_{\mu\in \mathbb{Y}_{n-1}}  L(\mu) \, \frac{v_{\mu+1}}{v_1^{1+|\mu+1|}},  \nn\\
%& t_1 = 1- \sum_{n=0}^\infty \frac{(-1)^n v^n}{n!} \sum_{\mu\in \mathbb{Y}_{n}}  L(\mu) \, \frac{v_{\mu+1}}{v_1^{1+|\mu+1|}}, \nn\\
& t_k = \delta_{k,1} - \sum_{n=0}^\infty \frac{(-1)^n v^n}{n!} \sum_{\mu\in \mathbb{Y}_{n+k-1}}  L(\mu) \, \frac{v_{\mu+1}}{v_1^{1+|\mu+1|}}. \label{tkandv}
\end{align}

\begin{prfn}{Lemma~\ref{thm-dlyz}} 
For $\gamma={\rm ch}_{2i_1-1}\cdots {\rm ch}_{2i_m-1}$ 
with $i_1,\dots,i_m\geq 1$, $m\geq0$ we have
\eqa
\HH_g(\gamma; \bt)
&=&\sum_{m_0,m_1,m_2,m_3,\dots \geq 0 \atop \sum_{i=0}^\infty (i-1)m_i =3g-3-\deg \gamma} 
\bigl\langle\gamma\, \tau_0^{m_0} \tau_1^{m_1} \tau_2^{m_2} \cdots \bigr\rangle_{g}\, \prod_{i=0}^\infty \frac{t_i^{m_i}}{m_i!}.\label{abc0}
\eeqa
($\gamma$ is defined as~1, if $m=0$.) 
Substituting the dilaton equation~\eqref{dilaton2} in~\eqref{abc0} we find
\eqa
\HH_g(\gamma; \bt) 
&=&\sum_{m_0,m_1,m_2,m_3,\dots \geq 0 \atop \sum_{i=0}^\infty (i-1)m_i =3g-3-\deg \gamma} 
\frac{(\sum_{i=0}^\infty m_i+2g-3)!}{(\sum_{i=0}^\infty m_i+2g-3-m_1)!} \, 
\langle\gamma\, \tau_0^{m_0}  \tau_2^{m_2} \tau_3^{m_3} \cdots \rangle_{g}\, \prod_{i=0}^\infty \frac{t_i^{m_i}}{m_i!}\nn\\
&=&\sum_{n=0}^\infty \frac{x^n}{n!} \sum_{m_2,m_3,m_4,\dots \geq 0 \atop \sum_{i=2}^\infty  (i-1)m_i =3g-3-\deg \gamma+n} \frac{\langle\gamma\, \tau_0^n\tau_2^{m_2}\tau_3^{m_3} \cdots \rangle_{g}}{(1-t_1)^{\sum_{i=2}^\infty m_i+2g-2+n}}\, \prod_{i=2}^\infty \frac{t_i^{m_i}}{m_i!}.\label{def2}
%&=&\sum_{n=0}^\infty \frac{x^n}{n!} \sum_{\lambda\in\mathbb{Y}_{3g-3-\deg\gamma+n}} 
%\frac{\langle\gamma\, \tau_0^n \tau_{\lambda+1} \rangle_{g}}{(1-t_1)^{\ell(\lambda)+2g-2+n}}\, \frac{t_{\lambda+1}}{m(\lambda)!}.\label{def2}
\eeqa
Substituting~\eqref{tkandv} in~\eqref{def2} and noticing that the dependence of~$t_1$ in $\HH_g(\gamma;\bt)$ is always through $1-t_1$, 
we find that there exists $\widetilde H_g(v,v_1,v_2,v_3,\dots)\in \CC[[v,v_1,v_2,v_3,\cdots,v_1^{-1}]]$ such that 
\[\HH_g(\gamma;\bt) = \widetilde H_g \bigl(v(\bt),v_1(\bt),v_2(\bt),\dots\bigr).\]
Then similarly to the proof of Lemma~\ref{ind-v-lem} we find that $\p \widetilde H_g/\p v=0$.  
Since the dependence on~$v$ of $\widetilde H_g(v,v_1,v_2,v_3,\dots)$ is {\it a priori} a power series, 
we can take $v=0$ when substituting~\eqref{tkandv} in~\eqref{def2}. 
So, if we associate to $v_m$ degree $m-1$ for $m\geq 1$, then $\widetilde H_g$ 
has the degree $3g-3-\deg \gamma$. Alternatively, 
if we assign $v_m$ another degree~$m$ for $m\geq 1$,  
then $\widetilde H_g$ has degree $2g-2$.  The lemma is proved. 
\end{prfn}

\end{appendix}

\bigskip

\noindent Boris Dubrovin

\noindent SISSA, via Bonomea 265, Trieste 34136, Italy

\noindent dubrovin@sissa.it

\medskip
\medskip

\noindent Di Yang

\noindent School of Mathematical Sciences, University of Science and Technology of China 

\noindent Hefei 230026, P.R. China 

\noindent diyang@ustc.edu.cn
\end{document}